\pdfoutput=1 
\documentclass[aps,twocolumn,prl,showpacs,amsmath,amssymb,10pt]{revtex4-1}
\usepackage{graphicx}
\usepackage{scrextend}
\usepackage{manfnt,pifont}
\usepackage{shuffle}

\usepackage{url}

\begin{document}


\title{Monodromy Relations for String Amplitudes on $AdS$}

\author{Luis F. Alday$^a$}
\author{Rodrigo S. Pitombo$^{a,b}$, Aurélie Strömholm Sangaré$^a$}

\affiliation{$^a$Mathematical Institute, University of Oxford, Andrew Wiles Building, Radcliffe Observatory Quarter, Woodstock Road, Oxford, OX2 6GG, U.K.}
\affiliation{$^b$ICTP South American Institute for Fundamental Research, Instituto de Física Teórica UNESP Rua Dr. Bento Teobaldo Ferraz 271, 01140-070, São Paulo, SP, Brazil}

\date{\today}

\begin{abstract}
\noindent 
We consider the $AdS$ Veneziano amplitude describing the scattering of four gluons in type IIB string theory on $AdS_5 \times S^3$. 
We propose a set of powerful monodromy relations between different colour-ordered amplitudes. These relations arise as a consequence of the emergent world-sheet description of open string scattering on $AdS$. In flat space, they reduce to the usual monodromy relations for the Veneziano amplitude, and they hold order by order in the small curvature expansion. The relations hold for all results available in the literature, including the scattering of arbitrary KK-modes.

\end{abstract}

\maketitle

\noindent{\bf Open string scattering on $AdS$.} For holographic $AdS$/CFT systems one can define string scattering on $AdS$ through correlators of local operators in the dual CFT.  Massless scattering amplitudes correspond to correlators of protected operators and the tree-level approximation corresponds to a large central charge limit \cite{Heemskerk:2009pn}. The study of gluon scattering on curved backgrounds through $AdS$/CFT was initiated in \cite{Alday:2021odx}. In the simplest set-up one adds probe branes to the standard $AdS$/CFT
system \cite{Aharony:1998xz,Ennes:2000fu,Karch:2002sh}. The scattering of open strings attached to these D-branes is dual to the correlator of currents
in the CFT. More specifically we focus on the set-up of \cite{Alday:2024yax,Alday:2024ksp}. In those papers the $AdS$ Veneziano amplitude for type IIB gluon scattering in $AdS_5 \times S^3$ was considered, and the first two curvature corrections around flat space were computed. The corresponding CFT is a 4d ${\cal N}=2$ superconformal field theory with $SU(2)_R \times U(1)_R$ R-symmetry and $SU(2)_L \times G_F$ flavour symmetry.  The $AdS$ Veneziano amplitude maps to the four-point correlator of ${\cal O}^I(x,v)$, the superconformal primary of the flavour multiplet. This is a Lorentz scalar operator, with protected dimension $\Delta=2$ and a singlet of $SU(2)_L$. Furthermore, it transforms in the adjoint of $SU(2)_R$, with polarisation vector $v^\alpha$, and the adjoint of $G_F$, with index $I$. The four-point correlator takes the form
\begin{eqnarray}
\langle {\cal O}^{I_1}(x_1,v_1) \cdots {\cal O}^{I_4}(x_4,v_4)  \rangle = \frac{v_{12}^2 v_{34}^2}{x_{12}^4 x_{34}^4}
G^{I_1 I_2 I_3 I_4}(z,\bar z,\alpha), \nonumber
\end{eqnarray}
where we have introduced the cross ratios 
\begin{equation}
\frac{x_{12}^2 x_{34}^2}{x_{13}^2 x_{24}^2}=|z|^2,~~~\frac{x_{14}^2 x_{23}^2}{x_{13}^2 x_{24}^2}=|1-z|^2,~~~\alpha=\frac{v_{13} v_{24}}{v_{12} v_{34}}, \nonumber
\end{equation}
with $x_{ij}=x_i-x_j$ and $v_{ij}= \epsilon_{\alpha \beta} v_i^\alpha v_j^\beta$. The solution to the superconformal Ward identities can be written in terms of a reduced correlator $H^{I_1 I_2 I_3 I_4}(U,V)$ such that
\begin{align*}
G^{I_1 I_2 I_3 I_4}(z,\bar z,\alpha)= \hspace{3pt}&G^{I_1 I_2 I_3 I_4}_0(z,\bar z,\alpha)\\
&+(1-z \alpha)(1-\bar z \alpha) H^{I_1 I_2 I_3 I_4}(z,\bar z), \nonumber
\end{align*}
where  $G^{I_1 I_2 I_3 I_4}_0(z,\bar z,\alpha)$ encodes the contribution from protected operators and  ${H}^{I_1 I_2 I_3 I_4}(z,\bar z)$ the non-trivial dynamics. From $H^{I_1 I_2 I_3 I_4}(z,\bar z)$ we define the reduced Mellin amplitude
\begin{eqnarray*}
H^{I_1 I_2 I_3 I_4}(z,\bar z)&=&\!  \int_{-i \infty}^{i \infty} \frac{ds dt}{(4\pi i)^2} |z|^{s+2}|1-z|^{t-2}\\
& & ~~~~~~~~~~~~~~~~~~~~\times \Gamma(s,t)M^{I_1 I_2 I_3 I_4}(s,t), \\
\Gamma(s,t)&=& \Gamma\left[1-\frac{s}{2} \right]^2 \Gamma\left[1-\frac{t}{2} \right]^2 \Gamma\left[1-\frac{u}{2} \right]^2,
\end{eqnarray*}
with $s+t+u=0$. In this letter we consider open string scattering where the world-sheet has the topology of a disk, with four insertions at its boundary. This corresponds to the leading non-trivial order in a large central charge expansion, while keeping the t'Hooft coupling constant $\lambda$ finite. At this order the colour structure of the correlator is given by the sum of single traces of products of generators $T^I$ of $G_F$, and we can decompose $M^{I_1 I_2 I_3 I_4}(s,t)$ into three colour-ordered amplitudes
\begin{align*}
&{ M}^{I_1 I_2 I_3 I_4}(s,t) =\text{Tr} \left( T^{I_1}T^{I_2}T^{I_3}T^{I_4} \right) {M}(s,t)\hspace{1pt} + \\
&\text{Tr} \left( T^{I_1}T^{I_4}T^{I_2}T^{I_3} \right) { M}(t,u) +\text{Tr} \left( T^{I_1}T^{I_3}T^{I_4}T^{I_2} \right) {M}(u,s).
\end{align*}
As shown, the three independent colour structures are related by crossing. In addition $M(s,t)=M(t,s)$. The $AdS$ Veneziano amplitude $A(S,T)$ is related to the colour-ordered Mellin amplitude $M(s,t)$ by the Borel transform
\begin{equation}
M(s,t) =\frac{8}{\lambda} \int_0^\infty d\beta e^{-\beta} \beta^3  A\left(\frac{2\beta s}{\sqrt{\lambda}}, \frac{2\beta t}{\sqrt{\lambda}}\right).
\end{equation}
From the perspective of scattering amplitudes on $AdS$ it is natural to use the $AdS$ radius $R$, related to  the t' Hooft coupling $\lambda$ at large central charge by $\sqrt{\lambda}=R^2/\alpha'$. The $AdS$ amplitude admits an expansion around flat space
\begin{equation}
A(S,T) = A^{(0)}(S,T) + \sum_{k=1}^\infty \left(\frac{\alpha'}{R^2}\right)^k A^{(k)}(S,T).
\end{equation}
At zeroth order we  recover the Veneziano amplitude in flat space
\begin{equation}
A^{(0)}(S,T)= \frac{1}{U} \int_0^1 x^{S-1}(1-x)^{T-1}dx,
\end{equation}
with $S+T+U=0$.
At higher orders in the curvature expansion the proposed structure is the following
\begin{equation}
A^{(k)}(S,T) = \frac{1}{U} \int_0^1x^{S-1}(1-x)^{T-1} g^{(k)}(S,T;x) dx,
\end{equation}
with $g^{(k)}(S,T;x)$ given by a linear combination of multiple polylogarithms (MPLs) $L_w(x)$, labelled by words in the alphabet $\{ 0,1\}$ up to length $3k$. More precisely
\begin{equation}
g^{(k)}(S,T;x)= \frac{1}{(S+T)^k} \sum_{n=0}^{3k} \sum_{|w|=n} P^{k}_w(S,T) L_w(x)
\end{equation}
with $P^{k}_w(S,T)$ homogeneous polynomials of degree $|w|$ (and labelled by both $w$ and $k$). MPLs $L_w(x)$ are recursively defined as follows. For the empty word $L_e(x)=1$. For a general word
\begin{eqnarray}
\frac{d}{dx} L_{aw}(x) = \frac{1}{x-a}  L_{w}(x),~~ ~a=0,1,
\end{eqnarray}
together with $L_w(0)=0$ for $w \neq 0^p$ and $L_{0^p}(x)=\frac{1}{p!} \log^p x$. The length of the word, denoted by $|w|$, is called the weight \footnote{Multiple zeta-values also have weight, since they are multiple polylogarithms evaluated at $x=1$. For instance, $\zeta(n)$ has weight $n$.}. Crossing symmetry implies
\begin{equation}
g^{(k)}(S,T;x)=g^{(k)}(T,S;1-x).
\end{equation}
The integrand is known for $k=0,1,2$, see \cite{Alday:2024yax,Alday:2024ksp}.

\vspace{0.2cm}
\noindent {\bf Monodromy relations in $AdS$.}  String theory colour-ordered disk amplitudes in flat space satisfy linear relations known as monodromy relations \cite{osti_4155317,Bjerrum-Bohr:2009ulz,Stieberger:2009hq}. For the Veneziano amplitude in flat space
\begin{equation}
e^{\pm i \pi S} A^{(0)}(S,U)+A^{(0)}(S,T)+e^{\mp i \pi T} A^{(0)}(U,T)=0,
\end{equation}
where recall $S+T+U=0$. In flat space, these relations can be derived directly from the world-sheet perspective. Indeed, consider the following integrals
\begin{equation}
\int_{{\cal C}^{\pm}} \eta(z)dz=0,~~~\eta(z)= z^{S-1}(1-z)^{T-1}
\end{equation}
where the contours ${\cal C}^{\pm}$ are given by straight lines just above/below the real line. The integrand $\eta(z)$ is a multi-valued function on the complex plane, with branch cuts along the real line for $x \leq 0$ and $x \geq 1$. We define it such that
  \begin{equation}
   \eta(x\pm i \epsilon) = \begin{cases}
        -e^{\pm i \pi S} (-x)^{S-1}(1-x)^{T-1} & \text{for $x<0$,} \\
        x^{S-1}(1-x)^{T-1} &\text{for $0<x<1$,} \\
        -e^{\mp i \pi T} (x)^{S-1}(x-1)^{T-1} &\text{for $x>1$}.
    \end{cases}
    \label{etacont}
\end{equation}
When integrating over the regions $(-\infty,0],[0,1]$ and $[1,\infty)$ we recover the three colour-ordered amplitudes $A^{(0)}(S,U),A^{(0)}(S,T),A^{(0)}(U,T)$ respectively, and the monodromy relations. Furthermore, note that different colour-ordered amplitudes correspond to different orderings of the vertex operators inserted at the boundary of the disk, so that for instance
\begin{eqnarray}
A^{(0)}(S,T)\sim \int_0^1dx  \langle V_1(0) V_2(x) V_3(1) V_4(\infty) \rangle.
\end{eqnarray}
The phases present in the monodromy relations can be seen as the result of exchanging the order of the vertex operators. Let us now analyse the situation in $AdS$. The phase $e^{\pm i \pi S}$ is directly related to the monodromy of $z^S$ as we move around the origin. The phase $e^{\mp i \pi T} $ is directly related to the monodromy of $(1-z)^T$ as we move around $z=1$. The presence of MPLs in $g^{(k)}(S,T;x)$ gives extra contributions as we move around zero and one. Let us denote by $K_0$ the operator such that the monodromy of MPLs around zero is given by the action of $e^{2\pi i K_0}$. Its action on MPLs is given by
\begin{equation}
K_0 L_{wa}(x) = L_w(x) \delta_{a0},~~~K_0 L_{e}(x) =0.
\end{equation}
In the same way, we introduce $K_1$ so that the monodromy around one is given by the action of $e^{2\pi i K_1}$. Its precise action on MPLs involves the Drinfeld associator and will be given below. We are now ready to write down the expected monodromy relations in $AdS$. Taking into account the monodromies of the prefactor $\eta(z)$ as well as the monodromies of the extra insertion $g(S,T;x)$ we obtain
\begin{equation}
e^{\pm i \pi (S+K_0)} A(S,U)+A(S,T)+e^{\mp i \pi (T+K_1)} A(U,T)=0.
\end{equation}
Several comments are in order. First, these relations are a very clear manifestation of - and were possible to guess thanks to - the world-sheet representation for the open string amplitude on $AdS$. Indeed, the action of $K_0,K_1$ is only defined at the level of the world-sheet integrand: they act non-trivially on MPLs, but trivially on rational functions of $S$ and $T$. Second, these proposed relations should hold order by order in the curvature expansion, and they do not mix different orders. Lastly, the action of $K_0,K_1$ on the flat-space result is trivial, and hence in this case we obtain the usual monodromy relations. We can also use crossing symmetry to write down the monodromy relations purely in terms of $K_0$
\begin{equation}
e^{\pm i \pi (S+K_0)} A(S,U)+A(S,T)+e^{\mp i \pi (T+K_0)} A(T,U)=0. \label{mon}
\end{equation}
Given the structure of the integrands, we can introduce the building blocks of the amplitude \cite{Alday:2025bjp}, a family of linear independent integrals labelled by words $w$
\begin{equation}
J_w(S,T) = \int_0^1 x^{S-1} (1-x)^{T-1} L_w(x) dx,
\end{equation}
 on which the operator $K_0$ acts as
\begin{equation}
K_0 J_{wa}(S,T) = J_w(S,T) \delta_{a0},~~~K_0 J_{e}(S,T) =0,
\end{equation}
 while leaving rational functions of $S$ and $T$ invariant. In order to understand the consequences of the monodromy relations (\ref{mon}), we need to work out the properties of the building blocks $J_{w}(S,T)$ under crossing of the $AdS$ Mandelstam variables. It is convenient to introduce non-commuting variables $(e_0,e_1)$ and a generating function for MPLs
\begin{equation}
L(e_0,e_1;x) = \sum_w L_w(x) e_w,
\end{equation}
with $e_w=1$ for the empty word and for instance $e_{0011}=e_0 e_0 e_1 e_1$. This induces a generating function for the $J-$integrals
\begin{equation}
{\cal J}(S,T;e_0,e_1)= \int_0^1 x^{S-1}(1-x)^{T-1} L(e_0,e_1;x)dx .
\end{equation}
Inspired by the monodromy relations in flat space, we will work out linear relations between ${\cal J}(S,T;e_0,e_1)$ and its crossing cousins. We start by considering the following integrals
\begin{equation}
\int_{{\cal C}^{\pm}} \eta(z)L(e_0,e_1;z) dz=0,
\end{equation}
where for ${\cal C}^{\pm}$ we have $z=x \pm i\epsilon$, just above/below the real line. MPLs have branch cuts along the real line for $x<0$ and $x>1$, so we have to be careful when doing contour manipulations. Splitting the contour of integration into three segments and making a simple change of variables we obtain
\begin{align*}
&{\cal J}(S,T;e_0,e_1) \\
&- e^{\pm i \pi S} \int_0^1  x^{-S-T}(1-x)^{S-1} L \left( e_0,e_1; 1 -\frac{1}{x} \pm i \epsilon \right)dx \\
&- e^{\mp i \pi T} \int_0^1 x^{-S-T}(1-x)^{T-1} L \left( e_0,e_1; \frac{1}{x} \pm i \epsilon \right)dx =0.
\end{align*}
We have used (\ref{etacont}) and changed  variables to bring the regions of integration to  $x \in [0,1]$ in all cases. Note that for $x \in [0,1]$ the generating functions in the second and third lines are evaluated just above/below the branch cuts. They can be related back to the generating function evaluated at $x$ by the identities \cite{vdH:fpsac99}
\begin{align*}
L \left( e_0,e_1; 1 -\frac{1}{x} \pm i \epsilon \right)&= L(-e_0-e_1,e_0;x)\\
&\hspace{8pt}\times Z(e_0,-e_0-e_1) e^{\pm i \pi e_0},\\
L \left( e_0,e_1; \frac{1}{x} \pm i \epsilon \right)&= L(-e_0-e_1,e_1;x)\\
&\hspace{8pt}\times Z(e_1,-e_0-e_1) e^{\mp i \pi e_1} Z(e_0,e_1),
\end{align*}
where the Drinfeld associator $Z(e_0,e_1)$ can be defined as the regularised generating function at $x=1$
\begin{equation}
Z(e_0,e_1) \equiv L(e_0,e_1;1) = 1 -\zeta(2)[e_0,e_1] +\cdots
\end{equation}
and satisfies $Z(e_0,e_1) Z(e_1,e_0)=1$. This leads to 
\begin{align}
{\cal J}(S,T;e_0,e_1) - {\cal J}(S,1-S-T;e_0,-e_0-e_1) e^{\pm i \pi (S + e_0 )} \nonumber \\
- {\cal J}(T,1-S-T;e_1,-e_0-e_1) e^{\mp i \pi (T + e_1 )} Z(e_0,e_1)=0, \label{relsunshifted}
\end{align}
where we have used
\begin{equation*}
{\cal J}(S,T;e_1,e_0)Z(e_0,e_1)={\cal J}(T,S;e_0,e_1).
\end{equation*}
Now combine (\ref{relsunshifted}) with the shift relations found in \cite{Alday:2025bjp}
\begin{eqnarray}
{\cal J}(S+1,T;e_0,e_1)=U_s {\cal J}(S,T;e_0,e_1)\\
{\cal J}(S,T+1;e_0,e_1)=U_t {\cal J}(S,T;e_0,e_1)
\end{eqnarray}
with $U_s=\frac{1}{S+T+e_0+e_1}(S+e_0),U_t=1-U_s$. This leads to 
\begin{align}
{\cal J}(S,T;e_0,e_1) -U_t^{-1} {\cal J}(S,U;e_0,-e_0-e_1) e^{\pm i \pi (S + e_0 )} \nonumber \\
- U_s^{-1} {\cal J}(T,U;e_1,-e_0-e_1) e^{\mp i \pi (T + e_1 )} Z(e_0,e_1)=0. \label{rels}
\end{align}
With this at hand, we can plug a given proposal for the amplitude into (\ref{mon}), and then use (\ref{rels}) to express the monodromy relations in terms of a set of conditions, each multiplying an independent building block $J_w(S,T)$. As we will see below, this gives highly non-trivial constraints on the amplitude. 

At the level of generating functions, $K_0$ and $K_1$ act as
\begin{gather*}
K_0 L(e_0,e_1;x)=L(e_0,e_1;x)e_0\\
K_1 L(e_0,e_1;x)=L(e_0,e_1;x)Z(e_1,e_0) e_1 Z(e_0,e_1)
\end{gather*}
and similarly when acting on the generating function ${\cal J}(S,T;e_0,e_1)$. We then see that the monodromies around $x=0$ and $x=1$, computed in \cite{FrancisB}, are given by
\begin{align*}
L(e_0,e_1;x + i \epsilon)&=e^{2\pi i K_0} L(e_0,e_1;x - i \epsilon),~~~x<0\\
L(e_0,e_1;x - i \epsilon)&=e^{2\pi i K_1} L(e_0,e_1;x + i \epsilon),~~~x>1.
\end{align*}

\vspace{0.2cm}
\noindent {\bf Implications.} 
Let us consider the first curvature correction to the Veneziano amplitude
\begin{equation}
A^{(1)}(S,T) = \sum_w \frac{P_w(S,T)}{U^2}J_w(S,T),
\end{equation}
where the sum runs over words of weight up to three, and $P_w(S,T)$ are homogeneous polynomials of degree $|w|$. The general ansatz consistent with crossing symmetry has 33 unfixed parameters. The monodromy relations give 28 independent linear constraints, leading to only five independent solutions! As expected, the amplitude found in \cite{Alday:2024yax} is among them. There is no solution with transcendentality one, see appendix. The simplest solution has transcendentality two:
\begin{align*}
&A^{(1)}(S,T) =\frac{J_e(S,T)-T J_0(S,T)-S J_1(S,T)}{U^2}\\
& \hspace{10pt}+\frac{S}{U} J_{00}(S,T)+\frac{T}{U} J_{11}(S,T)+J_{01}(S,T)+J_{10}(S,T).
\end{align*}
Finally, we also note that all five solutions satisfy the following constraints:
\begin{equation*}
\begin{gathered}
P_{01}(S,T)=P_{10}(S,T),\\
P_{110}(S,T)=P_{101}(S,T),~~~P_{001}(S,T)=P_{010}(S,T).
\end{gathered}
\end{equation*}
Namely, the MPLs that arise are the diagonal limit of single-valued MPLs. For more complicated examples we need to develop an efficient way to solve the monodromy relations. Introduce a linear operator $\Psi: e_w \to \mathbb{R}$ such that
\begin{equation}
A(S,T)=\langle \Psi(S,T;f_0,f_1)| {\cal J}(S,T;e_0,e_1) \rangle.
\end{equation}
This is achieved by defining 
\begin{equation}
\Psi(S,T;f_0,f_1) = \sum_w R_{\widetilde{w}}(S,T) f_w,
\end{equation}
with the notation $f_e=1,f_{001}=f_0f_0f_1$ and so on, together with an inner product in the space of words
\begin{equation}
\langle f_{\widetilde{w}}| e_{w'} \rangle = \delta_{w w'},
\end{equation}
where $\widetilde{w}$ is the operation that reverses the order of the letters. In this language we can write the monodromy relations as 
\begin{align}
&\langle \Psi(S,T;f_0,f_1)|{\cal J}(S,T;e_0,e_1)\rangle \nonumber\\
&+  \langle \Psi(S,U;f_0,f_1)|{\cal J}(S,U;e_0,e_1) e^{\pm i \pi(S+e_0)} \rangle\label{monlinear} \\
&+  \langle \Psi(U,T;f_0,f_1)|{\cal J}(T,U;e_1,e_0)e^{\mp i \pi(T+e_1)}  Z(e_0,e_1) \rangle = 0\nonumber.
\end{align}
The difference between the two linear relations (\ref{rels}) gives
\begin{align}
&{\cal J}(S,U;e_0,e_1) 
= K  {\cal J}(T,U;-e_0-e_1,e_1)  \label{difrels}\\
&~~ \times \sin \pi (T -e_0-e_1 )Z(e_0,-e_0-e_1) \left(\sin \pi (S + e_0 ) \right)^{-1}  \nonumber ,
\end{align}
with 
$$K= \frac{1}{S+T-e_1}(T-e_0-e_1 ) \frac{1}{S+e_0} (S+T-e_1).$$
Plugging (\ref{difrels}) into the difference of the two monodromy relations (\ref{monlinear}) we obtain
\begin{eqnarray*}
 \langle \Psi(S,U;f_0,f_1)|K  \tilde {\cal J}(T,U;-e_0-e_1,e_1) Z(e_0,-e_0-e_1) \rangle \\
=  \langle \Psi(U,T;f_0,f_1)|\tilde {\cal J}(T,U;e_1,e_0)Z(e_0,e_1) \rangle
\end{eqnarray*}
where $\tilde {\cal J}(T,U;e_1,e_0)={\cal J}(T,U;e_1,e_0)\sin \pi (T + e_1 )$. Using the properties for the inner product given in the appendix this turns into
\begin{eqnarray*}
\langle \Psi(S,U;-f_0,-f_0+f_1)|\tilde K  \tilde {\cal J}(T,U;e_0,e_1) Z(-e_0-e_1,e_0) \rangle \nonumber \\
= \langle \Psi(U,T;f_1,f_0)|\tilde {\cal J}(T,U;e_0,e_1)Z(e_1,e_0) \rangle
\end{eqnarray*}
with 
$$\tilde K= \frac{1}{ S+T-e_1}(T+e_0 ) \frac{1}{S-e_0-e_1} (S+T-e_1).$$
Since these equations are valid for any basis of independent functions $\tilde {\cal J}(T,U;e_0,e_1)$ we must have
\begin{align}\label{mr1}
\begin{split}
 Z(-\tau_0-\tau_1,\tau_0)  \Psi(S,U;&-f_0,-f_0+f_1) { \cal \tilde K} \\ 
 & = Z(\tau_1,\tau_0) \Psi(U,T;f_1,f_0),
\end{split}
\end{align}
where the operator $ {\cal \tilde K}$ acts from the right \footnote{The operators $\tau_i$ from the left/right act on $f_w$ as 
$$\tau_i f_{jw} = \delta_{ij} f_w,~~~ f_{wj} \tau_i = \delta_{ij} f_w.$$}
\begin{equation}
 { \cal \tilde K}= \frac{1}{S+T-\tau_1}(T+\tau_0 ) \frac{1}{S-\tau_0-\tau_1} (S+T-\tau_1).
\end{equation}
Combining crossing symmetry  $A(S,T)=A(T,S)$ with $ {\cal J}(T,S;e_0,e_1) =  {\cal J}(S,T;e_1,e_0) Z(e_0,e_1)$ we obtain
\begin{equation}
\Psi(S,T;f_0,f_1) = Z(\tau_1,\tau_0) \Psi(T,S;f_1,f_0), \label{cross}
\end{equation}
which combined with (\ref{mr1}) leads to 
\begin{align}\label{mreff}
\begin{split}
 Z(-\tau_0-\tau_1,\tau_0)  \Psi(S,U;-f_0,&-f_0+f_1)  {\cal \tilde K}\\
 &= \Psi(T,U;f_0,f_1).
 \end{split}
\end{align}
Equations (\ref{cross}) and (\ref{mreff}) do not make any reference to the integrals $J_w(S,T)$, but they are equivalent, of course, to crossing symmetry and the monodromy relations presented before. One can explicitly check that the first two curvature corrections found in \cite{Alday:2024yax,Alday:2024ksp}, $ \Psi^{(1)}(S,T;f_0,f_1) $ and $\Psi^{(2)}(S,T;f_0,f_1)$, satisfy these equations. At a given order in the small curvature expansion we have 
\begin{equation}
 \Psi^{(k)}(S,T;f_0,f_1) = \sum_{t=0}^{3k} \Psi^{(k)}_{t}(S,T;f_0,f_1) 
\end{equation}
where $\Psi^{(k)}_{t}(S,T;f_0,f_1)$ denotes the weight $t$ contribution. The maximally transcendental piece with $t=3k$ satisfies a simplified system of equations
\begin{align}\label{maxeqns}
\begin{split}
&\Psi^{(k)}_{3k}(S,T;f_0,f_1) = Z(\tau_1,\tau_0) \Psi^{(k)}_{3k}(T,S;f_1,f_0),\\
 &\frac{ \Psi^{(k)}_{3k}(S,U;-f_0,-f_0+f_1)}{S} \\
& \hspace{52pt}=Z(\tau_0,-\tau_0-\tau_1) \frac{\Psi^{(k)}_{3k}(T,U;f_0,f_1)}{T},
\end{split}
\end{align}
since any higher orders in the operator ${\cal \tilde K}$ would lower the weight. Consistency with the high-energy limit - large $S,T$ with $S/T$ fixed - implies
\begin{equation}
\Psi^{(k)}_{3k}(S,T;f_0,f_1)  = \frac{U^{k-1}}{k!} \left(\Psi^{(1)}_{3}(S,T;f_0,f_1)  \right)^{\shuffle,k} \label{max}
\end{equation}
where the notation $\left( \cdot \right)^{\shuffle,k}$ means shuffle product with itself, $k$ times.  As a result of the relations (\ref{Ronshuffle}) in the appendix, one can easily show that (\ref{max}) satisfies both equations in (\ref{maxeqns}) for general $k$, provided the result holds for $k=1$. Hence the monodromy relations are consistent with the high-energy limit, to all orders in the curvature expansion.    

\vspace{0.2cm}
\noindent{\bf Monodromy relations for general KK-modes.} The first curvature correction to the $AdS$ Veneziano amplitude corresponding to the scattering of arbitrary KK-modes on the sphere was constructed in \cite{Wang:2025owf}. The amplitude is labelled by the KK-modes $p_1,p_2,p_3,p_4$ and reduces to the one considered in \cite{Alday:2024yax,Alday:2024ksp} for $p_i=2$. The amplitude depends on the usual $AdS$ Mandelstam variables $S,T,U$, with $S+T+U=0$, and on the sphere Mandelstam variables $N_s,N_t,N_u$, with $N_s+N_t+N_u=-1$. For general KK-modes we expect the monodromy relations to take the form
 \begin{widetext}
\begin{equation}
e^{\pm i \pi (S+K_0)} A_{KK}(S,U,N_s,N_u)+A_{KK}(S,T,N_s,N_t)+e^{\mp i \pi (T+K_0)} A_{KK}(T,U,N_t,N_u)=0. \label{monkk}
\end{equation}
\end{widetext}
The full expression taken from \cite{Wang:2025owf} can be found in the appendix. One can explicitly check that it satisfies (\ref{monkk})!

\vspace{0.2cm}
\noindent{\bf Conclusions.} Over the last few years methods have been developed to compute tree-level string theory amplitudes on $AdS$ spaces \cite{Alday:2022uxp,Alday:2022xwz}. One of the most striking features of these results is the emergence of a world-sheet picture \cite{Alday:2023jdk,Alday:2023mvu}, both for closed as well as for open string amplitudes. In this letter we propose a set of monodromy relations for open string amplitudes in $AdS$. These arise as a consequence of the world-sheet description, are satisfied order by order in the small curvature expansion and in flat space they reduce to the usual monodromy relations for the Veneziano amplitude. The monodromy relations are very powerful in flat space, see \cite{Huang:2020nqy,Berman:2023jys,Chiang:2023quf}, and the same appears to be true in $AdS$. In the context of the proposal of \cite{Alday:2024yax,Alday:2024ksp}, and when combined with crossing symmetry alone, they reduce the number of free parameters from 33 to 5 for the first curvature correction, and from 565 to 86 for the second curvature correction. A natural direction is to use these new constraints to study and compute further curvature corrections. It would be interesting to understand the monodromy relations found in this letter in terms of braiding relations for open string vertex operators in $AdS$ backgrounds, along the lines of \cite{Boels:2010bv,Boels:2014dka}. More generally, it would be very interesting to understand how the structure studied in this letter arises from a first-principles world-sheet computation.  

\vspace{0.5cm}
\noindent{\bf Acknowledgements.} The work of L.F.A. is partially supported by the STFC grant ST/T000864/1. The work of R.S.P. is supported by the São Paulo Research Foundation (FAPESP) through the grants 2022/05236-1 and 2024/08232-2. For the purpose of open access, the authors have applied a CC BY public copyright licence to any Author Accepted Manuscript (AAM) version arising from this submission.

\appendix

\section{A toy model}

\noindent Let us consider a simple example where the amplitude is given by the sum of three terms up to weight one
\begin{equation}
A_{\text{toy}}(S,T) = \sum_{w=e,0,1} R_w(S,T)J_w(S,T),
\end{equation}
with $R_e(S,T),R_0(S,T)$ and $R_1(S,T)$ rational functions. Crossing symmetry under $S \leftrightarrow T$ implies
\begin{equation}
R_e(S,T)=R_e(T,S),~~~R_0(S,T)=R_1(T,S).
\end{equation}
In addition, the monodromy relations found in this letter imply the following non-trivial conditions
\begin{gather*}
S R_0(U,T)=T R_1(S,U),\\
T R_0(S,U)+T R_1(S,U) + S R_1(U,T)=0,\\
S T R_e(S,U)-S^2 R_e(U,T) = \\
S R_0(S,U)+ T R_0(S,U)+S R_1(S,U).
\end{gather*}
In the context of scattering amplitudes on $AdS$ we have
\begin{equation}
R_e(S,T)=\frac{\alpha}{(S+T)^{k+1}},~~R_0(S,T)=\frac{\beta S+\gamma T}{(S+T)^{k+1}}.
\end{equation}

For $k \neq 0$ we find no solutions. For $k=0$ we have the usual Veneziano amplitude together with
\begin{equation}
A_{\text{toy}}(S,T) = \frac{S J_0(S,T)+T J_1(S,T)}{U}.
\end{equation}

\section{Properties of the linear map}

\noindent In this letter we have defined the inner product
\begin{equation*}
\langle \Psi(f_0,f_1)| {\cal J}(e_0,e_1) \rangle = \sum_w R_w J_w,
\end{equation*}
with $\Psi(f_0,f_1) = \sum_w R_{\widetilde{w}} f_w, {\cal J}(e_0,e_1)=\sum_w J_w e_w$. This satisfies the following properties
\begin{align*}
\langle \Psi(f_0,f_1)| {\cal J}(e_0,e_1) &h(e_0,e_1) \rangle = \sum_{w,w'} R_{w w'} J_w h_{w'}\\
&= \langle h(\tau_0,\tau_1)  \Psi(f_0,f_1)| {\cal J}(e_0,e_1)  \rangle,
\end{align*}
with $h(e_0,e_1) = \sum_{w'} h_{w'} e_{w'}$. The operators $\tau_0,\tau_1$ act on a chain of $f'$ from the left as 
\begin{equation}
\tau_a f_{bw} = \delta_{ab} f_w.
\end{equation}
In the same way
\begin{multline*}
\langle \Psi(f_0,f_1)|  h(e_0,e_1){\cal J}(e_0,e_1) \rangle \\ 
=\langle \Psi(f_0,f_1) h(\tau_0,\tau_1) | {\cal J}(e_0,e_1)  \rangle.
\end{multline*}
The operators $\tau_0,\tau_1$ act on a chain of $f'$ from the right as 
\begin{equation}
 f_{wb}\tau_a = \delta_{ab} f_w.
\end{equation}
One can also prove the following relations
\begin{align*}
&\langle \Psi(f_0,f_1)| {\cal J}(e_1,e_0) \rangle=\langle \Psi(f_1,f_0)| {\cal J}(e_0,e_1) \rangle\\
& \langle \Psi(f_0,f_1)| {\cal J}(-e_0-e_1,e_1) \rangle\\
& \hspace{90pt}=\langle \Psi(-f_0,-f_0+f_1)| {\cal J}(e_0,e_1) \rangle.
\end{align*}
Another  property is the following. Let $\Psi_h(f_0,f_1)$ and $\Psi_g(f_0,f_1)$ be the two linear maps corresponding to integrands $h$ and $g$. Then, the linear map corresponding to their product is given by
\begin{equation}
\Psi_{h \times g}(f_0,f_1) = \Psi_h(f_0,f_1) \shuffle  \Psi_g(f_0,f_1),
\end{equation}
where the shuffle product acts on the space of non-commutative variables $f_0,f_1$ as
\begin{equation}
f_w \shuffle f_{w'} = f_{w \shuffle w'}.
\end{equation}
Now consider two linear maps 
\begin{equation}
\Psi^{(1)}(f_0,f_1) = \sum_w \psi^{(1)}_w f_w,~~~\Psi^{(2)}(f_0,f_1) = \sum_w \psi^{(2)}_w f_w,
\end{equation}
and an operator $R(\tau_0,\tau_1)=\sum \rho(w) \tau_w$, such that $\rho(w) \rho(w') = \rho(w \shuffle w')$. Examples of such an operator are $Z(\tau_0,\tau_1)$ and $Z(\tau_0,-\tau_0-\tau_1)$. Then
\begin{align}\label{Ronshuffle}
\begin{split}
&\left( R(\tau_0,\tau_1)\Psi^{(1)}(f_0,f_1) \right) \shuffle \left( R(\tau_0,\tau_1)\Psi^{(2)}(f_0,f_1) \right) \\
&=\sum_{w_1,w_2,w'_1,w'_2}  \rho(w_1 \shuffle w_2) \psi^{(1)}_{w_1 w'_1}\psi^{(2)}_{w_2 w'_2} \left( f_{w'_1} \shuffle f_{w'_2} \right)\\
&= \hspace{20pt}R(\tau_0,\tau_1) \left( \Psi^{(1)}(f_0,f_1)  \shuffle \Psi^{(2)}(f_0,f_1)  \right).
\end{split}
\end{align}
To see the last equality, focus on the specific term proportional to $\rho(z) f_{z'}$. This will receive contributions from all $w_1,w_1',w_2,w_2'$ such that $z \in w_1 \shuffle w_2$ and $z' \in w'_1 \shuffle w'_2$.

\section{$AdS$ Veneziano amplitude for KK-modes}
\noindent The first curvature correction to the $AdS$ Veneziano amplitude for general KK-modes was computed in \cite{Wang:2025owf}. It is given by
\begin{widetext}
\begin{align}
    A_{KK}^{(1)}&(S,T,N_s,N_t)=-J_e(S,T)\frac{6 N_s (\Sigma -3)+6 N_t (\Sigma -3)-5}{3 (S+T)^2}
    \nonumber\\
    &+\frac{(S (6 N_s (\Sigma -3)+2 \Sigma -3)+2 T (3 N_s (\Sigma -3)-\Sigma -1)) J_0(S,T)}{3 (S+T)^2}+\left(\frac{4 \Sigma }{3}-\frac{S^2-4 S T+T^2}{3 (S+T)^2}\right) J_{01}(S,T)
    \nonumber\\
    &+\frac{(2 S (3 N_t (\Sigma -3)-\Sigma -1)+T (6 N_t (\Sigma -3)+2 \Sigma -3)) J_1(S,T)}{3 (S+T)^2}+
    \left(\frac{4 \Sigma }{3}-\frac{S^2-4 S T+T^2}{3 (S+T)^2}\right) J_{10}(S,T)
    \nonumber \\
   &- \frac{2 S (2 \Sigma  (S+T)+S+4 T) J_{00}(S,T)}{3 (S+T)^2}-
    \frac{2 T (2 \Sigma  (S+T)+4 S+T) J_{11}(S,T)}{3 (S+T)^2}-J_e(S,T)\frac{3 S^3+3 S^2 T}{3 (S+T)^2}\zeta(3)
    \nonumber\\
   &
    +S \left(\frac{S}{S+T}+2\right) J_{001}(S,T)+S \left(\frac{S}{S+T}+2\right) J_{010}(S,T)+ \frac{T (2 S+3 T) J_{101}(S,T)}{S+T}+\frac{T (2 S+3 T) J_{110}(S,T)}{S+T}
    \nonumber\\&
         +\frac{2 S^2 J_{100}(S,T)}{S+T}-
    \frac{4 S^2 J_{000}(S,T)}{S+T}+
    \frac{2 T^2 J_{011}(S,T)}{S+T}-\frac{4 T^2 J_{111}(S,T)}{S+T}.\label{KKamp}
\end{align}
\end{widetext}
Here $S$ and $T$ are the usual $AdS$ Mandelstam variables, while $N_s$ and $N_t$ are the sphere Mandelstam variables, which are discretised. See \cite{Wang:2025owf} for details. Furthermore $\Sigma=(p_1+p_2+p_3+p_4)/2$. The scattering of the lowest KK-mode corresponds to $p_i=2$ and $N_s=N_t=-1/3$. 

\bibliography{refmom}
\bibliographystyle{utphys}

\end{document}